\documentstyle[11pt,newpasp,twoside]{article}
\def\edcomment#1{\iffalse\marginpar{\raggedright\sl#1\/}\else\relax\fi}
\reversemarginpar

\begin{document}
\title{Stellar population gradients in Seyfert 2 galaxies}
\author{Thaisa Storchi-Bergmann \& Daniel I. Raimann}
\affil{Instituto de F\'\i sica, UFRGS, Porto Alegre, Brasil}

\author{Rosa M. Gonz\'alez Delgado}
\affil{Instituto de Astrof\'\i sica de Andaluc\'\i a (CSIC), Granada, Spain}

\author{Henrique R. Schmitt\altaffilmark{1}}
\affil{National Radio Astronomy Observatories, Socorro, NM}

\author{Roberto Cid Fernandes}
\affil{Departamento de F\'{\i}sica, UFSC, Florian\'opolis, Brasil}

\author{Timothy Heckman}
\affil{Johns Hopkins University, Baltimore, MD}

\author{Claus Leitherer}
\affil{Space Telescope Science Institute, Baltimore, MD}

\altaffiltext{1}{Jansky Fellow}
 
\begin{abstract}

We study the variation of the stellar population properties
as a function of distance from the nucleus for a sample of
35 Seyfert 2 galaxies using the
technique of stellar population synthesis. We sample regions at the
galaxies with dimensions in the range
200$\times$200 to 400$\times$400\,pc and compare the synthesis results
with those of a control sample of non-Seyfert galaxies. We find that
both at the nucleus and up to 3 kpc from it the oldest age component
(10\,Gyr) presents a smaller contribution to the total flux
in the Seyfert than in the non-Seyfert galaxies of the same Hubble type,
while the components younger than 100\,Myr present a
larger contribution in the Seyfert's than in non-Seyferts.
In addition, while for the non-Seyferts clear gradients are present,
in which the contribution of the oldest components decrease with
distance from the nucleus and the contribution of the 1\,Gyr component
increases -- we do not find such gradients in most Seyferts.
These results suggest that the AGN-starburst connection
is a large scale phenomenon affecting not
only the inner few hundred parsecs, but the inner kiloparsecs.

\end{abstract}

\section{Introduction}

The present work has roots in the early and polemic work of
Roberto Terlevich and collaborators, in which they argued that starbursts
instead of mass accretion into supermassive black holes were the source
of energy of Active Galactic Nuclei (AGN; Terlevich \& Melnick 1985;
Terlevich, Diaz \& Terlevich 1990; Cid Fernandes \& Terlevich 1995).
By 1995, other groups would also find evidences
for the presence of starbursts around Seyfert nuclei (e. g. Heckman
et al. 1995, 1997; Gonz\'alez Delgado et al. 1998), or smaller
mass-to-light ratios than in non-Seyfert galaxies (Nelson \& Whittle 1996,
Oliva et al. 1999).

Our work in this subject comprises a number of papers in which we have investigated the
stellar population in  the nuclear region of these galaxies using the technique of
spectral synthesis, obtaining in particular
robust age indicators (Cid Fernandes et al. 1998;
Storchi-Bergmann et al. 1998; Schmitt et al. 1999; Storchi-Bergmann et al. 2000).
We have also investigated the correlation between the age of the stellar population
and with other properties, such as the infrared luminosity, galaxy morphology
and environment (Cid Fernandes et al. 2001a; Storchi-Bergmann et al. 2001).

In the present work, we discuss the results of two on-going studies
(Raimann et al. 2002, 2003), in which we investigate the radial
variation of the stellar population properties of a sample of
35 Seyfert 2 galaxies and compare them with those of a control
sample of non-Seyfert galaxies. The questions we want to answer are:
Which are the main differences between the stellar population of
Seyfert and non-Seyfert galaxies of the same Hubble type?
How far from the nucleus can we find this difference?
In Storchi-Bergmann et al. (2001) we have proposed an
evolutionary scenario in which  interactions are responsible
for triggering the activity and circumnuclear bursts of star formation.
If this is true, wouldn't the more external regions of the host galaxy also be affected?

\section{Sample and Data}

The Seyfert 2 sample
comprises two subsamples, which we call the northern (N) sample
(Gonz\'alez Delgado et al. 2001) and southern (S) sample
(Storchi-Bergmann et al. 2000). The N sample comprises the 20 brightest
Seyfert 2 galaxies with L([OIII]$>10^{40}$ergs\,s$^{-1}$
while the S sample comprises the
20 closest Seyferts 2 obeying the same luminosity criterium.

As a control sample of non-Seyfert galaxies we use the spectra of
10 nearby galaxies, distributed
in Hubble type as follows:
3 S0, 2 Sa's, 2 Sb's and 3 Sc's. In addition, we will also use as
a control sample the one of Bica (1988), comprising approximately
100 non-active galaxies spanning all Hubble types.

The data consists of long-slit optical spectra, which are
described in the papers listed above, except in the case of
Bica (1988) sample, which uses single  aperture data. Extraction of
one-dimensional spectra from the long-slit data were performed
at typical apertures of 2$\times$2 arcsec$^{2}$,
sampling regions at the galaxies of 200$\times$200\,pc$^2$ to
400$\times$400\,pc$^2$.
A sample of spectra at different distances from the
nucleus for one S0 Seyfert 2 is compared to that of a non-Seyfert S0 galaxy.
in Fig. 1.

\begin{figure*}
\vspace{9.0cm}
\caption{Sample nuclear and extranuclear spectra of the Seyfert 2 galaxy
Mrk\,1066 (left) and the non-Seyfert NGC\,6684 (right).}
\includegraphics{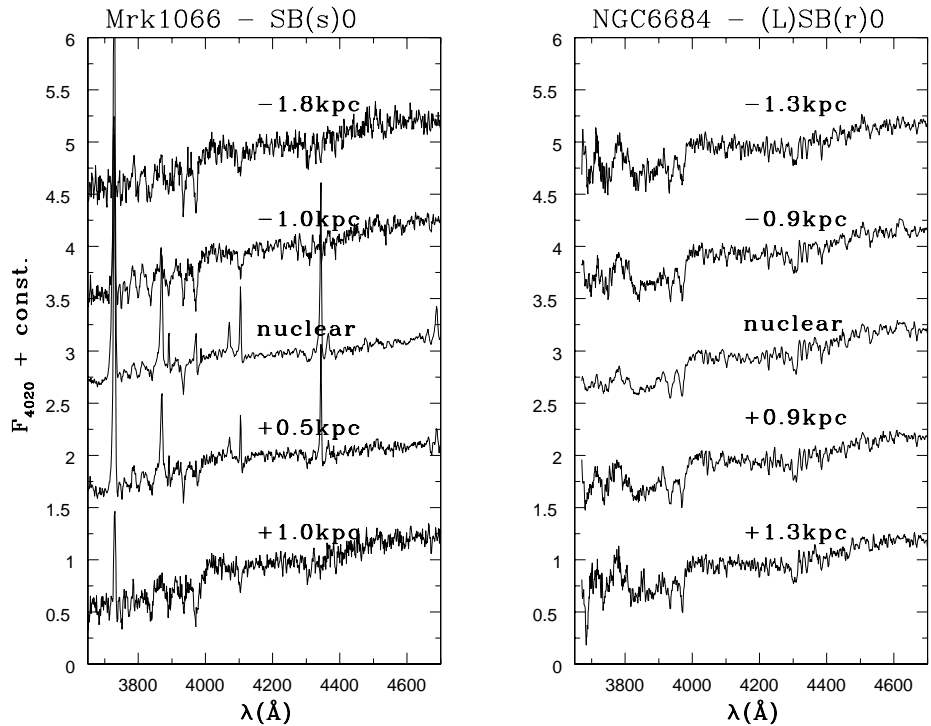}
\end{figure*}

\section{Measurements and Spectral Synthesis}

The measurements consist of
equivalent widths (hereafter W's) of six to eight absorption features
characteristic of the stellar population plus continuum flux ratios (hereafter C's)
at selected pivot points. The resulting values are illustrated in
Fig. 2 for one Seyfert and one non-Seyfert galaxy of Hubble type S0.

\begin{figure*}
\vspace{7.0cm}
\caption{Radial variations of the equivalent widths in \AA (Ws),
continuum color and surface brightness
(10$^{-15}$ erg cm$^{-2}$ s$^{-1}$ \AA $^{-1}$ arcsec$^{-2}$)
for a S0 non-Seyfert (left) and one Seyfert.
From top to bottom: W$_{wlb}$ (solid line) and W$_{H9}$
(dotted); W$_{CaIIK}$ (solid) and W$_{CaIIH+H\epsilon}$ (dotted);
W$_{CN-band}$ (solid) and W$_{G-band}$ (dotted). Vertical lines mark
distances of 1\,kpc and 3\,kpc from the nucleus.}
\label{var1}
\includegraphics{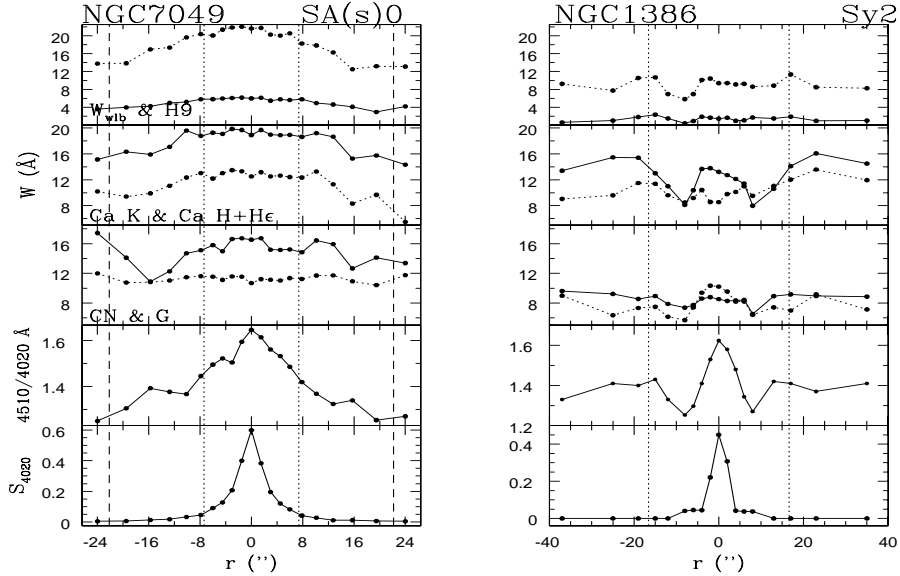}
\end{figure*}

The spectral synthesis was performed using the probabilistic formalism
described in Cid Fernandes et al. (2001a). We reproduce the observed W's
and C's using a base of star cluster spectra with
different ages and metallicities. We use 12 spectral
components representing the age-metallicity plane plus a 13$^{th}$
component representing a canonical AGN continuum
F$_{\nu} \propto \nu^{-1.5}$.

As we have pointed out in our previous works, in the
optical spectral range used in our work,
it is not possible to discriminate
between the FC and 3\,Myr old components for flux contributions
smaller than 40 per cent at 4020\AA,
because they have very similar continua. Therefore, in the
description of the synthesis results we combine the 3\,Myr
and FC components in one, the 3\,Myr/FC component.

These results are illustrated in Fig. 3 for one non-Seyfert and
one Seyfert galaxy of Hubble type S0,
where we compare the contribution of different age components to the
flux at $\lambda$4020\AA\ as a function of distance from the nucleus.
For the later Hubble types,
the gradients observed for the Sa-Sc non-Seyferts are similar to those observed for the
non-Seyfert S0 galaxies. Nevertheless, the
values for the contribution of the different age components to the
flux at $\lambda$4020\AA\ differ from those of the S0 galaxy. For example,
the nuclear contribution of the 10\,Gyr component in the case of the
S0 galaxies is larger than 80\%, while in the Sa-Sc control galaxies, it is around 60\%.
Consistently, the corresponding values for the contribution of
the 1\,Gyr component are smaller than 20\% for the former and reach
about 40\% for the latter. But the comparison between Seyfert
and non-Seyfert galaxies gives similar results to those illustrated in Fig. 3:
the contribution of the old age components is always smaller in
Seyfert's than in non-Seyferts.

\begin{figure}
\vspace{7.0cm}
\caption{Results of stellar population synthesis using a
base of star clusters spectra plus a power
law component F$_{\nu} \propto \nu^{-1.5}$ for a
non-Seyfert (left) and a Seyfert (right). The dots represent
percent contribution to the flux at 4020\AA\ from different age
components as a function of distance from the nucleus. The top panel shows
the internal reddening E(B-V)$_i$ obtained from the synthesis. Vertical
lines mark distances of 1 kpc and 3 kpc from the nucleus.}
\label{sin_normal}
\includegraphics{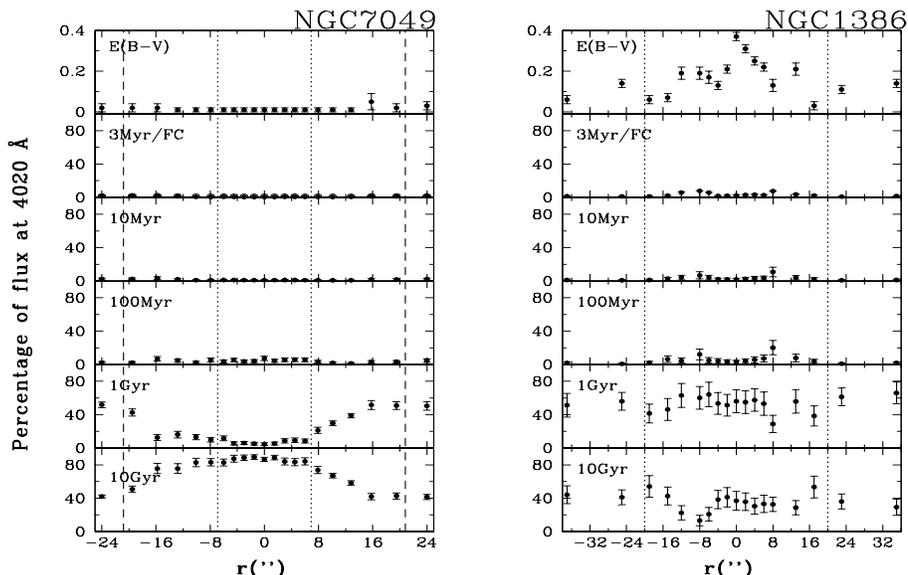}
\end{figure}


\section{Summary and Conclusions}

\begin{table*}
\tiny
\caption{Average contribution to the total flux at 4020\AA}
\label{average}
\begin{tabular}{lcccccccccccc} \hline
Hubble                 &  10G & & & & 1G  & & 100/10M & & & 3M/FC &  & \\
type                            & Nuc. & 1kpc & 3kpc & Nuc. & 1kpc & 3kpc & Nuc.
& 1kpc & 3kpc & Nuc. & 1kpc & 3kpc \\ \hline
S0 Sy 2                & 50 & 50 & -- & 26 & 37 & -- & 14 &  8 & -- & 10 &  5 &  -- \\
Non-Sy               & 86 & 77 & 50 &  6 & 15 & 36 &  9 &  8 & 12 &  0 &  0 &  1 \\
\\
Sa-Sb Sy 2             & 29 & 34 & 35 & 24 & 27 & 28 & 30 & 26 & 29 & 17 & 13 &  8 \\
Non-Sy            & 61 & (45) & -- & 33 & (52) & -- & 4 & (2) & -- & 2 & (1) & -- \\
Bica' Sa$^a$                & 56 & -- & -- & 37 & -- & -- &  4 & -- & -- &  2 & -- & --\\
Bica' Sb$^a$                & 53 & -- & -- & 35 & -- & -- &  7 & -- & -- &  5 & -- & -- \\
\\
Sbc Sy 2               & 31 & 25 & 17 & 25 & 26 & 32 & 33 & 24 & 32 & 12 & 26 & 19 \\
Non-Sy Sbc             & 68 & (43) & -- & 27 & (57) & -- & 3 & (0) & -- & 2 & (0) & -- \\
Non-Sy Sc              & 46 & -- & -- & 35 & -- & -- & 15 & -- & -- &  4 & -- & -- \\
Bica' Sc$^a$                & 27 & -- & -- & 27 & -- & -- & 26 & -- & -- & 20 & -- & -- \\
\\
S Sy 2                 & 31 & 30 & 34 & 25 & 29 & 32 & 24 & 24 & 24 & 20 & 16 & 10 \\ \hline
\end{tabular}

$^a$ Results for non-Seyfert galaxies from Bica (1988), in a region of 1\,kpc $\times$ 1\,kpc.
\end{table*}

The results of the synthesis are summarized in Table 1, which
shows that, for the Seyferts of all (S0-Sc) Hubble types,
the percent contribution of the oldest 10\,Gyr stellar component is
always smaller than in the control sample. This result
seems to hold up to at least 1 kpc from the nucleus and
in a few cases up to 3 kpc (in which the data with high
signal-to-noise ratio could be obtained that far from the nucleus).
Another difference  is in the contribution of the stellar
components younger than 100\,Myr, which are obviously larger in the
Seyfert than in the non-Seyferts, both at the nucleus and outside.

The only exception regards
the Sc Seyferts, when we compare the results with those of Bica (1988),
probably due to the
larger aperture of Bica's study, which may be including
the contribution of recent star-formation in the disk of the galaxies.

Regarding the radial variation of the contribution of the different
age components, we can
clearly see a gradient in the non-Seyfert
galaxies, which shows a decrease in the contribution
of the 10\,Gyr age component
and an increase of that of the 1\,Gyr component with distance from the nucleus
(the other components contributing with very small percentages).
This clear gradient observed in the non-Seyfert's is not
to present in most Seyfert galaxies.


The higher incidence of recent/intermediate age stellar populations in Seyfert
galaxies when compared with the control sample support an
AGN-star formation connection. Our results further suggest that,
in the evolutionary scenario proposed by Storchi-Bergmann et 
al. (2001), in which interactions are responsible for triggering
both star-formation and the AGN activity (which outlives the starburst),
the triggering of star-formation is not restricted to the nuclear
region but seems to extend to up to 3\,kpc from the nucleus.





\section{Discussion}

{\it Angeles Diaz:} How have you handled in your models the possible
presence of metallicity gradients through the inner parts of your
galaxies?

\bigskip

\noindent
{\it Thaisa Storchi-Bergmann:} The base of star cluster spectra contemplates
different metallicities, thus the effect is being taken into account. Nevertheless,
our experience has shown that the synthesis results are more robust
in terms of age, thus in the presentation of the results we add the contribution
of the components of different metallicities which have the same age.

\end{document}